# Tailoring the degree of entanglement of two coherently coupled quantum emitters

J.-B. Trebbia, Q. Deplano, P. Tamarat and B. Lounis

*1- Univ Bordeaux, LP2N, F- 33405 Talence, France*
*2- Institut d'Optique & CNRS, LP2N, F-33405 Talence, France*

**Abstract**

The control and manipulation of quantum-entangled non-local states is a crucial step for the development of quantum information processing. A promising route to achieve such states on a wide scale is to couple solid-state quantum emitters through their coherent dipole-dipole interactions. Entanglement in itself is challenging, as it requires both nanometric distances between emitters and nearly degenerate electronic transitions. Implementing hyperspectral imaging to identify pairs of coupled organic molecules trapped in a low temperature matrix, we reach distinctive spectral signatures of maximal molecular entanglement by tuning the optical resonances of the quantum emitters by Stark effect. We also demonstrate far-field selective excitation of the long-lived subradiant delocalized states with a laser field tailored in amplitude and phase. Interestingly, optical nanoscopy images of the entangled molecules unveil novel spatial signatures that result from quantum interferences in their excitation pathways and reveal the exact locations of each quantum emitter. Controlled molecular entanglement can serve as a test-bench to decipher more complex physical or biological mechanisms governed by the coherent coupling and paves the way towards the realization of new quantum information processing platforms.

The manipulation and coherent control of arrays of interacting quantum systems are at the heart of the second quantum revolution. Solid-state quantum emitters such as single molecules[1], quantum dots[2,3] and color centers in diamond[4] are promising candidates for the realization of such arrays in quantum networks[5] as they can easily be manipulated with light and integrated in quantum photonics devices[6,7]. Entangling the electronic states of coherently interacting solid-state emitters is challenging since it requires both a coupling strength larger than the coherence decay rate, implying nanometric distances between the emitters, and quasi-degenerate optical transitions, detuned by less than their coupling strength. Among the candidates to achieve this goal, quantum dot molecules[8] are formed by coherent tunneling between individual quantum dots separated by a controlled nanometric distance. Yet, in addition to their complexity and cost of fabrication, these materials suffer from a severe drawback inherent to their self-assembling growth procedure: The poor control over size, strain and composition of the constituent quantum dots, results in large detuning of their emission energies[8].

Polycyclic aromatic hydrocarbon molecules embedded in well-chosen solid matrices at liquid helium temperatures have proven to be test-bench systems for quantum optics[9-13], and are intrinsically identical. Although promising, the production of entangled molecular electronic states for quantum technologies[14] remains hindered by the difficulty to fulfill the requirements of space and transition-frequency proximities of the molecules, due to their random spatial distribution in the host crystal and the variety of local matrix environments that span their optical resonances within an inhomogeneous band. Seeking two molecules matching these coincidences within in a solid matrix is like looking for a needle in a haystack. Coherent optical dipole coupling of a single pair of molecules has been reported only once over the past two decades, in an experiment combining near-field scanning probe microscopy and spectroscopy[15]. Indeed, such local probe techniques are limited in scalability and heavy to implement in cryogenic environments. Further challenges are to manipulate the degree of entanglement of pairs of molecules having frozen geometries and dipole orientations, and selectively address any quantum entangled state. For instance, engineering subradiant states is promising for quantum information storage with real time processing[16-21]. The associated line narrowing may also improve the sensitivity to external fields for quantum metrology applications[22,23].

Here, using a hyperspectral imaging approach to unveil coupled pairs of fluorescent molecules in highly doped molecular crystals, we demonstrate the manipulation of their degree of entanglement through Stark shifts of their optical resonances. We achieve maximal entanglement, where the single-excitation eigenstates of the coupled pair - the subradiant and the superradiant states - are nearly pure Bell states[24]. Delocalized molecular electronic states are found to extend over distances as large as 60 nm, which opens up attractive perspectives in terms of addressability of the quantum emitters. Interestingly, far-field super-resolution optical nanoscopy[25] images of the entangled molecules reveal novel spatial features that result from quantum interferences in their excitation pathways and reveal the exact locations of each molecule. Furthermore, we develop a method of coherent and selective excitation of the long-lived subradiant state, which paves the way to the realization of new quantum information schemes based on subradiance[26-28].

The host-guest system chosen in the experiment consists of dibenzanthanthrene (DBATT) molecules randomly embedded in a naphthalene film with a thickness less than few micrometers. At 2 K, thermal dephasing of their optical coherences vanishes, so that single molecules exhibit sharp zero-phonon lines (ZPL) with a natural linewidth of the order of 20 MHz set by the inverse lifetime of the excited state[29,30]. Such molecules nearly behave like simple two-level systems[31] and present a remarkable photostability, allowing quantum optical measurements at strong excitation intensities[32]. The sample is placed at the top of a gold inter-digitated electrode comb with a 20 µm spacing period in order to apply external electric fields up to the MV.m$^{-1}$ range and thus tune the molecular optical resonances by Stark effect[33] (see Fig. 1a and Supplementary Fig. 1). In the search for coupled molecules, hyperspectral fluorescence microscopy is performed while the laser frequency is scanned across the ~3 nm-wide inhomogeneous band centred on 618 nm. We then select pairs of spatially unresolved molecules having resonance frequencies within the continuous scan range of our laser (< 10 GHz). An optical saturation study is then performed on such pairs to seek the spectral signature of dipole-dipole coupling, which manifests in the excitation spectrum as the onset of a central resonance with increasing excitation intensities, as exemplified in Fig. 1b. This central resonance is located at the exact center of the two ZPLs present in the low intensity spectrum and its peak amplitude has a quadratic dependence on the excitation intensity[15]. Such evolution is the hallmark of two-photon simultaneous excitation of both molecules in the state $|E\rangle = |e_1, e_2\rangle$ from the ground state $|G\rangle = |g_1, g_2\rangle$, where $|g_i\rangle$ and $|e_i\rangle$ ($i$ = 1,2) are the ground and first electronic excited states of the bare molecules, respectively (Fig. 1c). Using this method, we have identified a dozen pairs of coupled molecules with various dipole-dipole configurations and coupling strengths. The characteristic spectral fingerprints of three of them are presented in the fluorescence excitation spectra of Fig. 1d,e,f, other examples being displayed in Supplementary Fig. 2. Figure 1e recalls the so-called J-aggregate configuration[34], where the orientations of the transition dipoles $\hat{d}_1, \hat{d}_2$ are aligned along the molecular separation direction $\hat{r}_{12}$, while Fig. 1f evokes the so-called H-aggregate configuration with parallel dipoles nearly orthogonal to $\hat{r}_{12}$.

The coupling strength of the interacting molecules, separated by a distance $r_{12}$ much smaller than their transition wavelength, is given by[35,36] $V = 3\alpha\gamma_0[(\hat{d}_1 \cdot \hat{d}_2) - 3(\hat{d}_1 \cdot \hat{r}_{12})(\hat{d}_2 \cdot \hat{r}_{12})]/4(kr_{12})^3$. In this expression, $\gamma_0$ is the radiative decay rate of the electronic excited states $|e_1\rangle$ and $|e_2\rangle$, $k = n(\omega_1 + \omega_2)/2c = n\omega_0/c$ is the wavenumber associated to the mean value $\omega_0$ of the molecular resonance frequencies $\omega_1$ and $\omega_2$ (detuned by $\Delta = \omega_1 - \omega_2$) in a medium with refraction index $n$, and $\alpha$ is the combined Debye-Waller/Franck-Condon factor of the molecules representing the fraction of photons emitted on the purely electronic transition. Strong dipole-dipole coupling $|V| > \gamma_0$ can be reached for separation distances as large as tens of nm with nearly parallel transition dipoles and lead to the formation of two entangled delocalized states: A symmetric, superradiant state $|S\rangle = a|g_1, e_2\rangle + b|e_1, g_2\rangle$ with enhanced radiative decay rate $\gamma_+ = \gamma_0 + 2ab\gamma_{12}$ and an antisymmetric, subradiant state $|A\rangle = a|e_1, g_2\rangle - b|g_1, e_2\rangle$ with reduced radiative decay rate $\gamma_- = \gamma_0 - 2ab\gamma_{12}$ (Fig. 1c). The cross-damping rate $\gamma_{12} = \alpha\gamma_0(\hat{d}_1 \cdot \hat{d}_2)$ arises from the incoherent coupling of the bare molecules through the vacuum field[36]. The states $|S\rangle$ and $|A\rangle$ are split by $2\Lambda = \sqrt{\Delta^2 + 4V^2}$ and have projections $a = \sqrt{(\Delta/2 + \Lambda)^2/((\Delta/2 + \Lambda)^2 + V^2)}$ and $b = \sqrt{V^2/((\Delta/2 + \Lambda)^2 + V^2)}$ on the bare states[37], which provide a measure of the degree of molecular entanglement. Maximally entangled states (Bell states) corresponding to $a = b = 1/\sqrt{2}$ can be reached when the detuning between the bare molecules is such that $|\Delta| \ll |V|$. Since fluorescent molecules are trapped in the host matrix with fixed positions and orientations that set their coupling strength $V$, their degree of entanglement can be tuned only by adjusting their resonance frequencies.

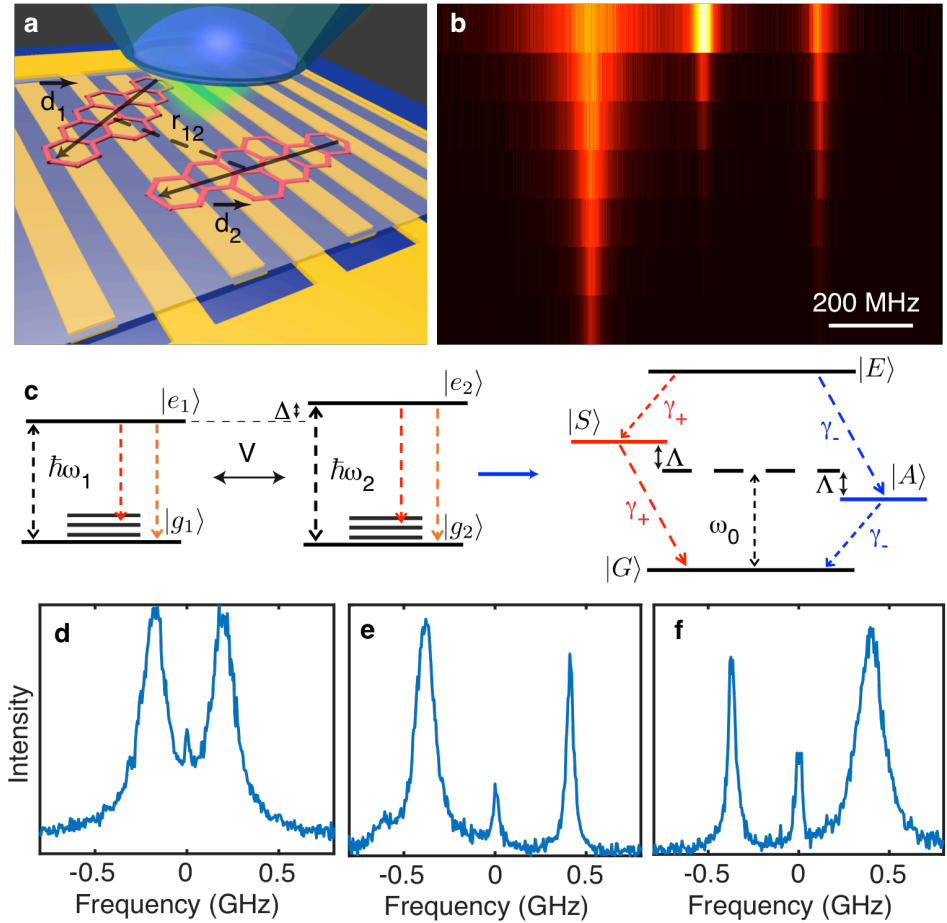

**Figure 1 : Spectral signatures of coherently coupled pairs of molecules.**
**a**, Schematic of the sample. A highly doped naphthalene crystal doped with DBATT molecules is placed above a comb of micro-patterned gold electrodes (period of 20 μm) to apply electric fields up to several MV m$^{-1}$. A microscope objective with numerical aperture 0.95 is used to excite the molecules on their ZPLs and collect the red-shifted emitted photons. **b**, Series of fluorescence excitation spectra recorded for a pair of coupled molecules at different excitation intensities ranging from 2 W cm$^{-2}$ to 250 W cm$^{-2}$. The onset of a two-photon transition shows direct evidence of coherent dipole-dipole interaction. **c**, Left: Energy levels of two bare molecules. Besides the purely electronic transition (green arrow), the vibrational levels may be involved in the fluorescence process (red arrows) but do not participate to the coherent coupling due to fast relaxation to the ground electronic state (on the picosecond timescale). Right: The coupled molecules are treated as a four-level system with a ground state $|G\rangle$, a doubly excited state $|E\rangle$ and two split entangled states $|A\rangle$ and $|S\rangle$. **d-f**, Excitation spectra of three different pairs of coupled molecules with various coupling strengths and dipole orientations.

The manipulation of entanglement by Stark shifting the molecular transitions with a static electric field is demonstrated in Fig. 2 and Supplementary Fig. 3. The spectrum in Fig. 2a presents the two ZPLs of a coupled molecular system (in the J-configuration) consisting of two largely detuned molecules ($|\Delta| \gg |V|$). These lines displaying similar widths and comparable amplitudes arise from the excitation of nearly localized excited molecular states $|e_1, g_2\rangle$ and $|g_1, e_2\rangle$ ($ab\sim 0$ and $\gamma_+ \sim \gamma_- \sim \gamma_0$). When decreasing $|\Delta|$ (Fig 2b), the linewidth of the lower-energy, superradiant level broadens while that of the higher-energy, subradiant level sharpens down to 13 MHz. This linewidth is significantly smaller than the natural linewidth ~ 23 MHz of uncoupled DBATT molecules embedded in naphthalene[29] (Supplementary Fig. 4) and shows direct evidence for molecular subradiance. Since $\alpha \sim 0.35$ for such molecules (supplementary Fig. 5), this result points to a radiative relaxation rate $\gamma_- \sim \gamma_0(1-\alpha)$ of a subradiant state formed by two nearly parallel molecular dipoles in the nearly maximal entanglement regime. Indeed, the rate $\gamma_-$ reaches its lower bound, as the hallmark of a complete suppression of fluorescence on the purely electronic transition. Interestingly, the reduced energy splitting between the two entangled states obtained at the minimal detuning $|\Delta|$ is a signature of a weak dipole-dipole coupling ($V \sim \gamma_0$), which means that the coherent interaction can create delocalized entangled states between solid-state quantum emitters separated by a distance as large as ~ 60 nm.

Maximal entanglement is also demonstrated for a pair of molecules having nearly parallel transition dipoles in the H-configuration. Figures 2c,d present the fluorescence excitation spectra of their subradiant and superradiant ZPLs recorded in the low saturation regime. The linewidths of these lines $\gamma_-/2\pi \sim (13.9 \pm 0.6)$ MHz and $\gamma_+/2\pi \sim (33.0 \pm 1.6)$ MHz are consistent with their lower and upper bounds $\gamma_\pm \sim \gamma_0(1 \pm \alpha)$. The saturation plots of the ZPL widths presented in Fig 2e are well reproduced with similar natural linewidths. This confirms the ability to reach the complete delocalization of the excitation over the coupled molecules with maximal superradiance and subradiance effects. Voltage manipulation of the entanglement is presented in Fig. 2f in the high excitation regime. At large molecular detunings (highest negative voltages), two ZPLs broadened by saturation are observed together with a dim central line assigned to two-photon excitation of $|E\rangle$. As the detuning is reduced, the central line becomes more pronounced and a clear anti-crossing develops between the $|S\rangle$ and $|A\rangle$ levels at the highest positive voltage. The fluorescence intensity of the lowest-energy line vanishes, which reflects the formation of a subradiant antisymmetric state $|A\rangle$ that becomes less efficiently excited with a Gaussian-shaped laser beam centered between the molecules. In contrast, the ZPL associated to the symmetric state $|S\rangle$ broadens due to a more efficient laser excitation.

In order to reproduce these behaviors and characterize the maximal degree of entanglement achieved by Stark effect, we compute the fluorescence signal proportional to the sum of the stationary populations of the states $|e_1, g_2\rangle$, $|g_1, e_2\rangle$ and twice that of $|E\rangle$. These populations are derived from the solutions of the Lindblad master equation governed by the Hamiltonian $H$ written in the rotating wave approximation as[37]

$$H = \hbar\omega_0(|E\rangle\langle E| - |G\rangle\langle G|) + \hbar\Lambda(|S\rangle\langle S| - |A\rangle\langle A|)$$
$$-\frac{\hbar}{2}\{[(a\Omega_1 + b\Omega_2)|E\rangle\langle S| + (b\Omega_1 + a\Omega_2)|S\rangle\langle G|]e^{i\omega_L t}$$
$$+ [(a\Omega_2 - b\Omega_1)|E\rangle\langle A| - (b\Omega_2 - a\Omega_1)|A\rangle\langle G|]e^{i\omega_L t} + h.c.\}$$

(Eq. 1)

where the first two terms describe the coupled molecular system, while the third is its interaction with a driving laser field of frequency $\omega_L$. The Rabi frequencies are defined as $\Omega_i = -\vec{d}_i \cdot \vec{\mathcal{E}}(\vec{r}_i - \vec{r}_c)/\hbar$, where $\vec{d}_i$ (i = 1,2) are the molecular transition dipoles and $\vec{\mathcal{E}}(\vec{r}_i - \vec{r}_c)$ the complex amplitudes of the laser field at the locations $\vec{r}_i$ of the molecules with respect to the laser spot center $\vec{r}_c$. The Hamiltonian matrix elements that govern the laser coupling of the ground state $|G\rangle$ with $|A\rangle$ and $|S\rangle$ are respectively $\langle A|H|G\rangle \propto a\Omega_1 - b\Omega_2$ and $\langle S|H|G\rangle \propto b\Omega_1 + a\Omega_2$. Their dependence on $\Delta$ provide a straightforward understanding of the spectral trails in Figs. 2f,g. For nearly parallel molecular dipoles, a Gaussian-shaped beam centered at mid-distance between the molecules will generate in-phase dipole oscillations with $\Omega_1 = \Omega_2$. When $\Delta$ approaches zero ($a \approx b \approx 1/\sqrt{2}$), the excitation of the symmetric (superradiant) state is therefore enhanced while that of the antisymmetric (subradiant) state becomes forbidden. The simulated spectral trails shown in Fig. 2g quantitatively reproduce the experimental trajectories of Fig. 2f in positions, widths and intensities, considering $V = 10\,\gamma_0$. At the limit of the voltage breakdown for our electrode array (~150 V), the molecular detuning is brought down to $\Delta = 5\,\gamma_0$ and nearly optimal entanglement is achieved with $a = 1.12/\sqrt{2}$ and $b = 0.86/\sqrt{2}$.

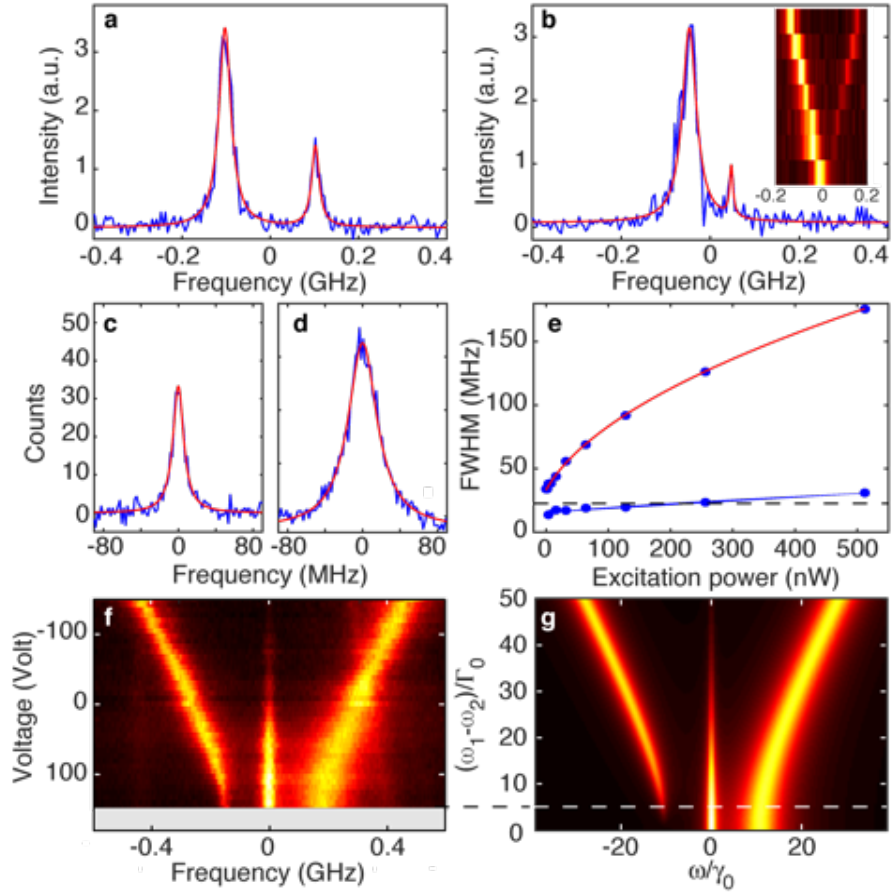

**Figure 2. Manipulation of the molecular entanglement by Stark effect.**
**a,b**, Excitation spectra of a pair of molecules with nearly parallel dipoles in the J-configuration and $V \sim \gamma_0$, recorded with an excitation intensity of 0.25 W cm$^{-2}$. The applied electric field shifts the molecular detuning from $10\,\gamma_0$ (**a**) to $4\,\gamma_0$ (**b**). Lorentzian fits (red curves) of the subradiant (higher-energy) ZPL gives a FWHM linewidth of 22 MHz in **a** and 13 MHz in **b**. The inset in **b** displays the spectral trails of this pair as the voltage is swept from 0 (bottom) to 150 V (top). **c,d**, ZPLs of the subradiant (**c**) and the superradiant (**d**) states of another molecular pair with nearly parallel dipoles in the H-dipole configuration and $V \sim 15\,\gamma_0$. The excitation powers are respectively 0.1 nW and 0.3 nW. The red curves are Lorentzian fits with FWHM linewidths $(13.9 \pm 0.6)$ MHz and $(33.0 \pm 1.6)$ MHz, respectively. **e**, Saturation plots (blue disks) and fits (red curves) of these ZPL linewidths. Radiative rates $\gamma_-/2\pi = 15.3 \pm 0.5$ MHz and $\gamma_+/2\pi = (33.0 \pm 0.5)$ MHz are derived from the fitting saturation intensities. The horizontal dashed line at 23 MHz marks the average linewidth of single DBATT molecules at low excitation intensities. **f,g**, Experimental (**f**) and simulated (**g**) Stark-tuned spectral trails of a coupled pair of molecules under an excitation intensity of 200 W cm$^{-2}$. The applied voltage is limited to the range -150 V- 150 V by electrode breakdown. As the system is tuned towards the anti-crossing point, the two-photon transition gains weight, the superradiant ZPL widens and the subradiant ZPL narrows with vanishing fluorescence amplitude. The simulations are performed for a H-dipole configuration leading to maximal entanglement at the anti-crossing point.

Striking differences in the excitation efficiencies of $|S\rangle$ and $|A\rangle$ also manifest in the fluorescence intensity autocorrelation function $g^{(2)}(\tau)$ recorded under resonant excitation of $|S\rangle$ and $|A\rangle$ at identical laser intensities. Indeed, besides strong photon antibunching on both transitions, the normalized coincidence histograms show Rabi oscillations that are much faster on the $|G\rangle \rightarrow |S\rangle$ transition (Fig. 3a) than on the $|G\rangle \rightarrow |A\rangle$ one (Fig. 3b). Photon bunching[15] is evidenced when the laser is tuned to the central two-photon resonance (Fig. 3c), as a signature of simultaneous excitation of coherently coupled molecules to the doubly excited state $|E\rangle$ with the subsequent two-photon emission cascade (Fig. 3c). These behaviors are markedly different from the autocorrelation expected for two uncoupled molecules, which would display photon antibunching with a value $g^{(2)}(0) = 0.5$ at zero-delay time. These correlation functions are well reproduced with numerical simulations, using molecular and laser interaction parameters deduced from prior analysis of the fluorescence excitation spectra, as well as polarization data (Supplementary Note 1 and Supplementary Figs. 6,7).

Selective preparation of long-lived entangled states is key for the realization of many quantum information schemes[27,28] and quantum memories[26]. Considering a parallel dipole configuration, shaping the laser field with identical amplitudes and opposite phases at the molecular positions ($\Omega_1 = -\Omega_2$) will enable the excitation of the subradiant state and forbid that of the symmetric superradiant state. We show here that the excitation of the sole antisymmetric subradiant state can be achieved, using a circularly polarized doughnut-shaped (first order Laguerre-Gaussian) beam whose zero-field center is placed midway between both emitters. In the configuration of nearly maximal entanglement, selective excitation of either the superradiant or the subradiant state is theoretically (Fig. 3d) and experimentally (Fig. 3e) demonstrated, using alternatively Gaussian and doughnut types of excitations, respectively. Indeed, the ZPL associated to the $|S\rangle$ state is barely observable in the case of a doughnut beam excitation, while it dominates the $|A\rangle$ line when exciting the system with a Gaussian beam (Fig. 3e). This simple far-field approach is a direct illustration of the ability to tailor selection rules with laser field sculpting. Moreover, it does not suffer from disturbances and quenching effects occurring when the emitters interact with non-uniform near-fields of metallic nanostructures[38].

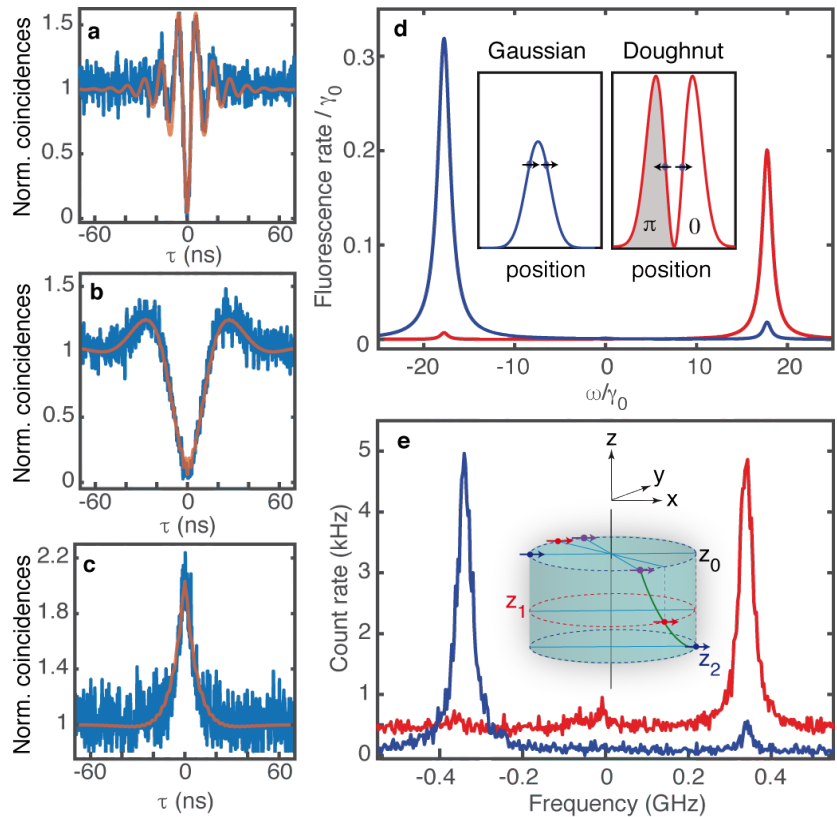

**Figure 3. Selective excitation of superradiant and subradiant states.**

**a,b,c**, Normalized photon coincidence histograms (after background subtraction) measured for a pair of coupled molecules excited with a Gaussian-shaped laser beam of intensity 200 W cm$^{-2}$ at resonance with the symmetric state (**a**), the antisymmetric state (**b**) and the two-photon transition (**c**). They are characterized by an antibunching dip and Rabi oscillations in (**a**) and (**b**), while photon bunching is evidenced in (**c**) as a signature of two-photon emission. The second-order correlation functions (red curves) are computed for nearly parallel dipoles in the J-configuration, $V = -17 \gamma_0, \gamma_{12} = 0.3 \gamma_0$, and $\Delta = 10 \gamma_0$ (Supplementary Note 1). Simulated (**d**) and experimental (**e**) spectra of the same pair excited with a Gaussian-shaped beam of 8 nW (blue curves) and a doughnut-shaped beam of 1 µW (red curves) centered midway between the molecules. The associated profiles of the laser field amplitude and phase are schematically displayed in the inset of (**d**). The simulations are performed with the above values of $V, \gamma_{12}, \Delta$, and with $a = 0.8, b = 0.6$. Depending on whether the emitters are excited in phase (Gaussian beam) or out of phase (doughnut beam), exclusive pumping of the superradiant or the superradiant state, respectively, is performed. The inset of (**e**) exemplifies three parallel-dipole configurations represented by blue, red, and purple couples of arrows, leading to the same coupling strength $V$ and spectra (Supplementary Fig. 8). The positions of the dipoles may differ along the optical (z) axis, while their projections in the focal (x,y) plane are separated by 18 nm.

While a complete characterization of the molecular and laser interaction parameters is required for a thorough investigation of the entanglement signatures, only partial information can be extracted from the above spectroscopic measurements. Indeed, the inset of Fig. 3e illustrates configurations of parallel dipoles with different positions that all lead to reproduce the experimental spectra (Supplementary Fig. 8). Far-field super-resolution imaging techniques allowing nanometric localization of single emitters could be a tool of choice to locate interacting quantum emitters at the nanometer scale. The simplest approach, based on successive localizations of fluorescence spot centers as the laser frequency is tuned[39], will fail at super-localizing the positions of entangled emitters, since $|S\rangle$ and $|A\rangle$ are spatially delocalized over both molecules. We show in Fig. 4 that the Excited-State Saturation (ESSat) nanoscopy technique[25,40] can reveal the positions of nanometrically spaced coupled emitters through subtle spatial signatures of delocalized quantum states. Using this method, fluorescence images are recorded by scanning over the coupled pair (Fig. 4a) a tightly focused doughnut beam with frequency $\omega_L = (\omega_1 + \omega_2)/2$. At weak saturations, the fluorescence image essentially reflects the intensity distribution of the doughnut-shaped laser beam, and a single fluorescence dip is observed (inset of Fig. 4a). For two *uncoupled* molecules with identical resonance frequencies, the saturation of the molecular transitions would broaden the fluorescence spot with the onset of two sharp dips at the exact positions of the molecules (Supplementary Fig. 9). However, for two coherently coupled molecules subject to different complex Rabi frequencies $\Omega_1 = \Omega(\vec{r}_1 - \vec{r}_c)$ and $\Omega_2 = \Omega(\vec{r}_2 - \vec{r}_c)$, non-intuitive features take shape in the ESSat image. Indeed, as the excitation intensity is raised, the central dark spot splits into two fluorescence dips at fixed positions (Fig 4b). Moreover, two additional dips develop in the direction orthogonal to that of the first ones, with a depth and separation that depend on the intensity (Fig 4c). Simulations of the fluorescence signals taking into account the vectorial nature of the field at the focal spot[41] reproduce well the experimental ESSat images (Fig. 4b,c) for a pair of highly entangled molecules with parallel transition dipoles. They provide a unique mean to determine the locations of the emitters with a nanometric precision. Indeed, it turns out that none of the fluorescence dips coincides with the actual locations of the molecules, which are located at the extremities of the segment equal and orthogonal to the one connecting the fixed two dips.

In order to interpret these striking image structures on the nanometric scale, we calculate the spatial dependence of eigenenergies of the molecular system coupled to the laser field and the populations of $|G\rangle$, $|S\rangle$, $|A\rangle$ and $|E\rangle$ (Supplementary Note 2). When the doughnut-shaped laser spot is scanned along the inter-molecular x-axis, the fluorescence drops when two dressed levels become degenerate (Fig. 4d). Such coincidences are accompanied by a population increase for $|G\rangle$ and $|A\rangle$ (Fig. 4e). The number of fluorescence dips as well their positions strongly depend on the excitation intensity (Supplementary Note 3 and Supplementary Fig. 10). When the spot is scanned along the perpendicular (y-axis) direction, both emitters undergo the same laser intensity ($|\Omega_1| = |\Omega_2|$) and the fluorescence dips occur at the positions where the phases taken by the laser field at the molecular positions differ by $\pi/2$. Such situations translate into an increase in the population of $|G\rangle$ at the expense of that of the excited states, as shown in Fig. 4f. This can be interpreted as the result of a destructive interference between the two excitation paths which bring the system from $|G\rangle$ to $|E\rangle$: $|G\rangle \rightarrow |A\rangle \rightarrow |E\rangle$ and $|G\rangle \rightarrow |S\rangle \rightarrow |E\rangle$. In the quasi-resonant regime, the two-photon transition can be described by an effective interaction Hamiltonian with a coupling matrix element[42] $\hbar\Omega_{EG}^{eff} = \frac{\langle E|H|A\rangle\langle A|H|G\rangle}{\hbar(\omega_0-\omega+\Lambda)} + \frac{\langle E|H|S\rangle\langle S|H|G\rangle}{\hbar(\omega_0-\omega-\Lambda)}$. At resonance $\omega = \omega_0$, one can readily see from the matrix elements of $H$ that $\Omega_{EG}^{eff}$ vanishes at the two fixed spot positions where $\Omega_2 = \Omega_1 e^{\pm i\frac{\pi}{2}}$. Overall, besides valuable information collected on the geometric configuration of the coupled emitters, the structuration of the laser field in amplitude and phase enables the inception of novel methods to selectively address quantum-entangled states and build quantum interference pathways among them.

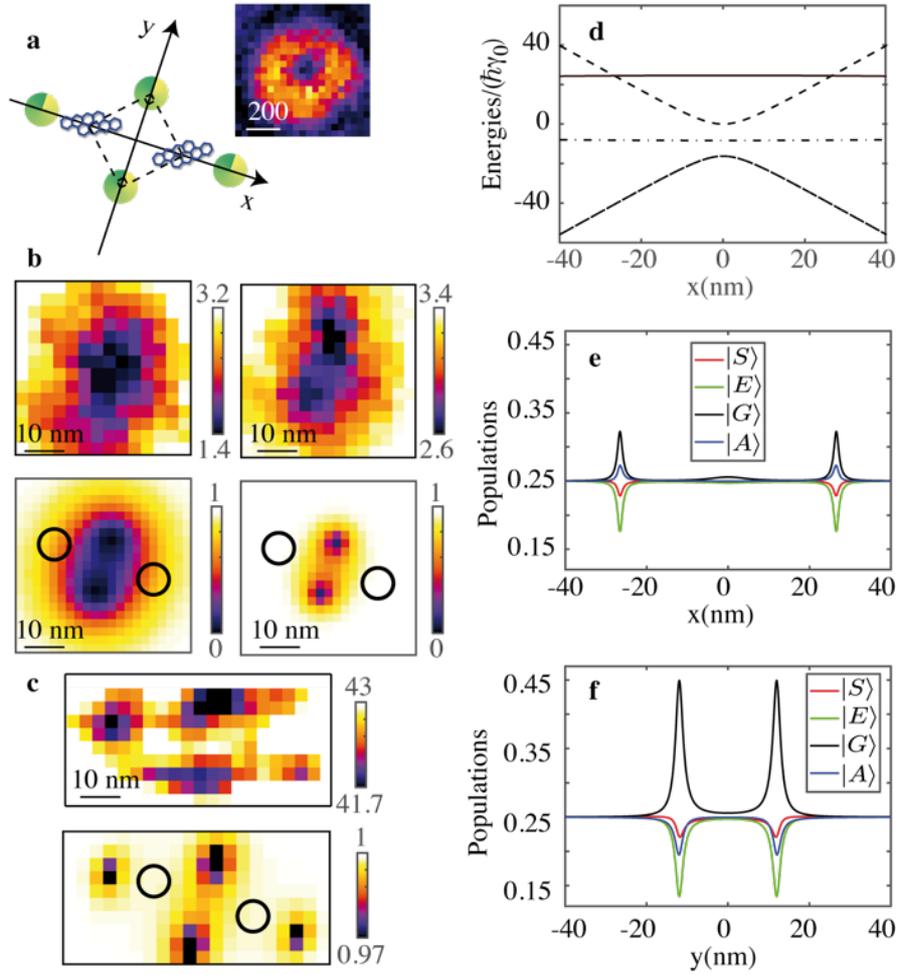

**Figure 4 : Optical nanoscopy of molecules undergoing coherent optical dipole-dipole coupling.**
**a**, Configuration of the molecules in the focal plane (x,y). The green disks mark the four positions of the doughnut beam center leading to fluorescence dips. For the two positions marked along the y-axis, the phase difference between the laser fields felt by the two molecules is $\pi/2$. Inset: Fluorescence image recorded at weak saturations. **b,c**, Experimental (top frames) and simulated ESSat (bottom frames) images of a pair of coupled molecules excited with a laser tuned on the two-photon transition, with increasing doughnut-peak intensities: 125 W cm$^{-2}$, 500 W cm$^{-2}$ (**b**), $5.4 \times 10^3$ W cm$^{-2}$ (**c**). The simulated ESSat images are computed with $\Delta = 0, V = -22\,\gamma_0, \gamma_{12} = 0.5\,\gamma_0$ and intensities of 175 W cm$^{-2}$, 700 W cm$^{-2}$ (**b**), $7 \times 10^3$ W cm$^{-2}$ (**c**). The in-plane positions of the emitters, separated by (22 ± 5) nm, are marked with a circle. **d**, Evolution of the four eigenenergies of $H$ as the excitation beam is swept along the x-axis. **e**, Evolution of the stationary populations of $|G\rangle, |A\rangle, |S\rangle$ and $|E\rangle$ along the x-axis. Each eigenstate experiences a position-dependent light shift and the fluorescence signal drops at the degeneracy points. Such coincidences are accompanied by a population increase for the $|G\rangle$ and $|A\rangle$. **f**, Evolution of the stationary populations of $|G\rangle, |A\rangle, |S\rangle$ and $|E\rangle$ as the laser spot is swept along the y-axis. The peaks of $|G\rangle$ population occur at the two positions where $\Omega_2 = \Omega_1 e^{\pm i\frac{\pi}{2}}$.

In summary, we demonstrate the manipulation of the degree of entanglement of coherently coupled solid-state quantum emitters, using Stark shifts of their electronic levels. Nearly maximal quantum entanglement is achieved for emitters separated by distances up to several tens of nanometers in various dipole configurations. Moreover, a simple far-field optical method is developed to selectively excite the super- and subradiant states, which opens new opportunities in quantum information processes, in particular those exploiting the long-lived radiative decay of subradiant states. Novel spatial signatures inherent to dipole-dipole interaction are unveiled with ESSat nanoscopy and provide a unique mean to localize the coupled emitters. This far-field versatile method could be extended to engineer the spectroscopic selection rules for various molecular or nanometer scale systems, or to demonstrate fast manipulation of entanglement for quantum logic gate operations[43]. The experimental and theoretical tools developed here to study the test-bench entangled molecular pair should also foster thorough investigations of a wealth of elaborate coupled systems, such as polymer conjugates[44] and quantum dot molecules[8]. A particularly interesting extension of this work will be to study the evolution of the entanglement within a system coupled to a controlled thermal bath, with a view to exploring possible non-Markovian processes and collective quantum effects responsible for the efficiencies of light-harvesting complexes[45].